# An Effective Round Robin Algorithm using Min-Max Dispersion Measure

Sanjaya Kumar Panda
Department of Computer Science & Engineering
National Institute of Technology, Rourkela
Rourkela, India
sanjayauce@gmail.com

Sourav Kumar Bhoi
Department of Computer Science & Engineering
National Institute of Technology, Rourkela
Rourkela, India
souravbhoi@gmail.com

***Abstract*— Round Robin (RR) scheduling algorithm is a preemptive scheduling algorithm. It is designed especially for time sharing Operating System (OS). In RR scheduling algorithm the CPU switches between the processes when the static Time Quantum (TQ) expires. RR scheduling algorithm is considered as the most widely used scheduling algorithm in research because the TQ is equally shared among the processes. In this paper a newly proposed variant of RR algorithm called Min-Max Round Robin (MMRR) scheduling algorithm is presented. The idea of this MMRR is to make the TQ repeatedly adjusted using Min-Max dispersion measure in accordance with remaining CPU burst time. Our experimental analysis shows that MMRR performs much better than RR algorithm in terms of average turnaround time, average waiting time and number of context switches.**

***Keywords-Operating System, Round Robin, Min-Max Round Robin, Turnaround time, Waiting time, Context switch.***

## I. INTRODUCTION

An Operating System is a collection of programs and utilities. It is an interface between end user and system hardware, so that the user can handle the system in a convenient manner [5]. Proportional share resource management provides a flexible and useful abstraction for multiplexing time shared resources [2]. Modern Operating Systems become more complex, they have evolved from a single task to a multitasking environment in which processes run in a concurrent manner [2] [3]. CPU scheduling algorithms decides which of the processes in the Ready Queue (RQ) is to be allocated to the CPU. There are many different CPU scheduling algorithms, out of those algorithms, Round Robin (RR) is the oldest, simplest and most widely used proportional share scheduling algorithm [1] [2]. It is similar to FCFS scheduling, but preemption is added to switch between processes [4]. A small unit of time is used in RR which is called as Time Quantum (TQ) or Time Slice (TS). The CPU scheduler goes around the RQ, allocating the CPU to each process for a time interval of up to 1 TQ [4]. If new process is arrived then it is added to the tail of the circular queue. The CPU scheduler picks the first process from the queue, sets a timer to interrupt after one TQ and dispatches the process [1]. After TQ is expired, the CPU preempts the process and the process is added to the tail of the circular queue. If process finishes before the end of the TQ, the process itself preempts the CPU willingly. In this paper, we present a solution to the TQ problem by adjusting TQ with respect to the existed set of processes in RQ.

## II. PRELIMINARIES

CPU scheduling is the basis of multi programmed operating system. The idea of multiprogramming is, if a process is waiting for an I/O request, then the CPU switches from that process to another process. So, the CPU is always busy in multiprogramming. But in a simple computer system, the CPU is idle until the I/O request is finished. All computer resources are scheduled before in use. So, CPU scheduling algorithm determines how the CPU will be allocated to the process. CPU scheduling algorithms are two types, one is non-preemptive and another is preemptive scheduling algorithms. In non-preemptive scheduling, once the CPU is allocated to a process , the process keeps the CPU until it releases the CPU either by terminating or by switching to the waiting state. But, in preemptive scheduling, the CPU can release the processes even in the middle of the execution. A process is a program at the time of execution. A process is more than the program code; it includes the program counter, the process stack, and the contents of process register etc. A process is a dynamic object. The processes are assigned to a processor are put in a queue called Ready Queue. CPU Utilization is the percentage of time that the processor is busy. It generally ranges from 0 to 100 percent. Throughput means how many processes are finished by the CPU with in a time period. The time interval between the submission of the process and time of the completion is the Turnaround time. Waiting time is the amount of time a process is waiting in the RQ, waiting in I/O and waiting in CPU. The number of times CPU switches from one process to another is called as the number of context switches. There are well known CPU scheduling algorithms that has been developed such as First Come First Serve (FCFS) algorithm, Shortest Job First (SJF) algorithm, Shortest Remaining Time Next (SRTN) algorithm, Round Robin (RR) algorithm and Priority Scheduling algorithm. RR and SRTN are preemptive in nature. RR is most suitable for time sharing systems. But its average output parameters (turn-around time, waiting time, etc.) are not feasible enough to be employed in real-time systems.

## III. RELATED WORK

In last few years different approaches are used to increase the performance of RR scheduling in different ways. Basically, the CPU scheduler is concerned mainly with CPU utilization, throughput, turnaround time, waiting time, response time and fairness [3]. Self-Adjustment Time Quantum in Round Robin (SARR) [1] algorithm is based on a new approach called dynamic TQ, in which TQ is repeatedly adjusted according to the current burst time of the running processes. Dynamic Quantum with Re-adjusted Round Robin (DQRRR) [5] algorithm is based on a TQ, in which TQ is calculated as median of the existed set of processes. A. Bhunia [6] has proposed an approach to increase performance of Multi Level Feedback Scheduling (MLFQ) in which response time of starved processes and over all turnaround time of the whole scheduling process decreases around eight to ten percent.

## IV. PROPOSED APPROACH

In this approach, time quantum is taken as the range of the CPU burst time of all the processes. The range of the processes is the difference between the largest (maximum) and smallest (minimum) values.

### *A. Uniqueness of Our Approach*

Let's assume that the data are sorted in increasing numerical order. It gives better turnaround time and waiting time. Generally, the performance of RR algorithm depends upon the size of static Time Quantum (TQ). If the TQ is extremely large, the algorithm approximate to First-Come First-Served (FCFS). If the TQ is extremely small, the algorithm causes too many context switches. So, our approach solves this problem by taking a dynamic TQ where the TQ is the difference between maximum and minimum CPU burst time as shown in equation (1).

$$TQ = MAXBT - MINBT \qquad (1)$$

Where MAXBT = MAXimum Burst Time

MINBT = MINimum Burst Time

### *B. Proposed Algorithm*

In our algorithm, processes are already present in the Ready Queue (RQ). By default, Arrival Time (AT) is assigned to zero. The number of processes 'n' and CPU Burst Time (BT) are accepted as input and Average Turnaround Time (ATT), Average Waiting Time (AWT) and number of Context Switch (CS) are produced as output. Let TQ and $TQ_{new}$ be the time quantum and new time quantum respectively. The pseudocode for the algorithm is presented in Figure 1 and the flowchart of the algorithm is presented in Figure 2.

```
1. All the processes present in the ready queue are sorted in ascending order.
   //n = number of processes, i = loop variable

2. while ( RQ != NULL )
    //RQ = Ready Queue
    TQ = MAXBT – MINBT
    //TQ = Time Quantum
   //MAXBT = MAXimum Burst Time
   //MINBT = MINimum Burst Time
   (Remaining burst time of the processes)
   // If one process is there then TQ is equal to BT of itself

3. if (TQ < 25)
           set TQnew = 25
   else
           set TQnew = TQ
   end if

4. //Assign TQ to (1 to n) process
   for i = 1 to n
       {
           Pi → TQnew
       }
   end for
    // Assign TQnew to all the available processes.

5.  Calculate the remaining burst time of the processes.

6. if ( new process is arrived and BT != 0 )
            go to step 1
   else if  ( new process is not arrived and BT != 0  )
            go to step 2
   else if  ( new process is arrived  and  BT  == 0)
            go to step 1
   else
            go to step 7
   end if
   end while

7. Calculate ATT, AWT and CS.
   //ATT = Average Turnaround Time
   //AWT = Average Waiting Time
   //CS = number of Context Switches

8. End
```

Figure 1. Pseudo code for Min-Max Round Robin (MMRR) algorithm.

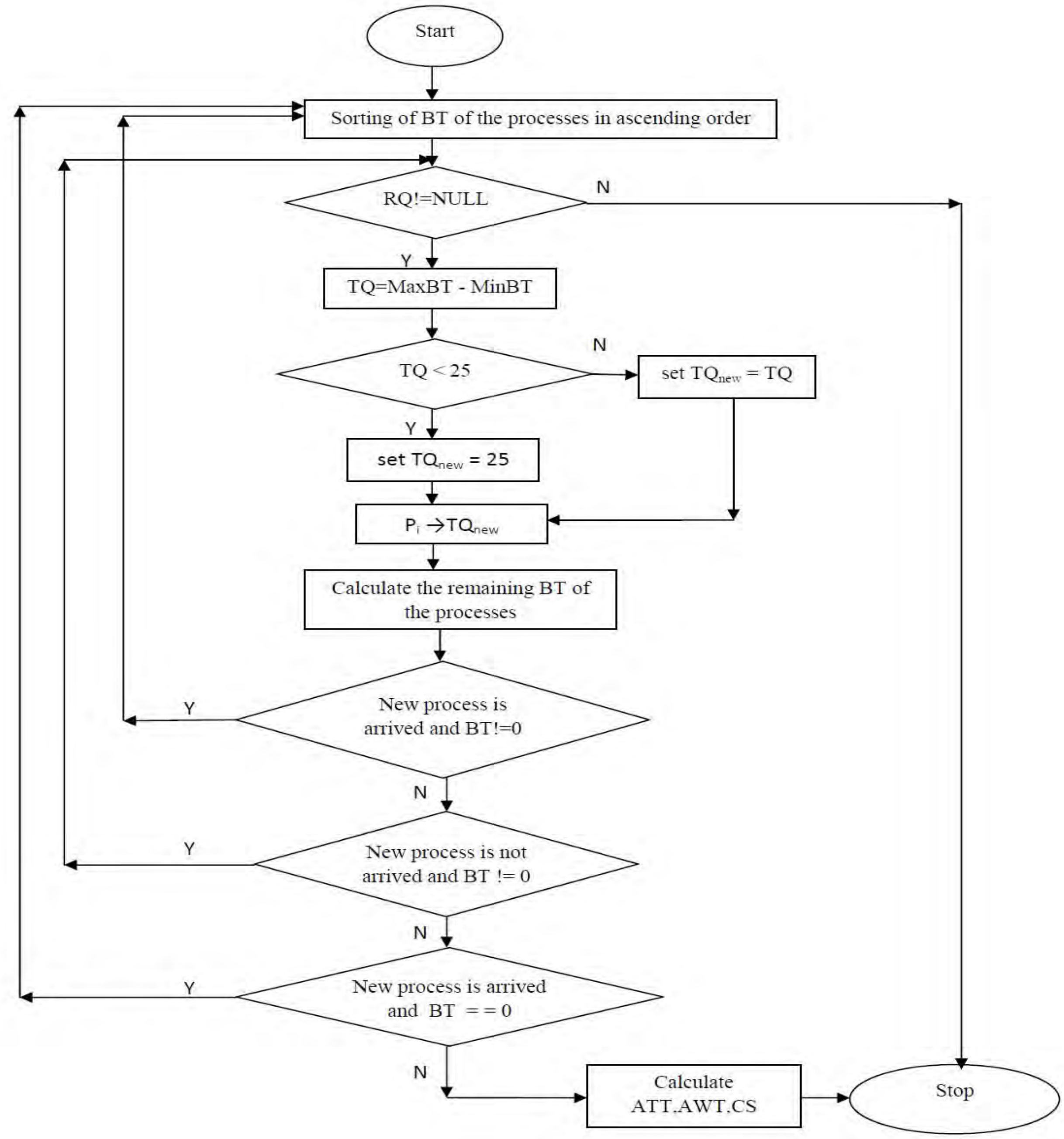

Figure 2. Flowchart of Min-Max Round Robin (MMRR) algorithm

### *C. Illustration*

Suppose four processes arriving at time = 0, and CPU burst time sequence P1 = 90, P2 = 96, P3 = 9, P4 = 37. The processes are sorted in ascending order of their CPU burst time which results in sequence P3 = 9, P4 = 37, P1 = 90, P2 = 96. Then TQ is calculated. TQ is the difference between maximum CPU burst time i.e. P2 = 96 and minimum CPU burst time i.e. P3 = 9. So, TQ is equal to 96 – 9 = 87. After first iteration the remaining CPU burst time sequence is P3 = 0, P4 = 0, P1 = 3, P2 = 9. In this case, the processes P3 and P4 are deleted from the Ready Queue (RQ). Again, CPU burst time is sorted in ascending order and new TQ is calculated. Here, new TQ is equal to 9 – 3 = 6 which is less than 25. So, new TQ is set to 25. After second iteration the remaining CPU burst time sequence is P1 = 0, P2 = 0. P1 and P2 are deleted from the RQ. Since, no process in the RQ, it completes its execution and ATT, AWT and CS are calculated. In this case, ATT = 127.5, AWT = 69.5, CS = 5.

## V. EXPERIMENTAL ANALYSIS

### *A. Assumptions Taken*

The system environment where all the experiments are performed is a single processor environment and all the processes are independent. Assume that all processes are CPU bounded. Time Quantum (TQ) is not more than maximum burst time. The processes are sorted in ascending order of their CPU burst time. We assume a constant TQ equal to 20 in all cases. The context switching time is equal to zero i.e. there is no context switch overhead incurred in transferring from one process to another. Let us assume that M represents the Min-Max TQ. If the Min-Max TQ is less than 25 then its value must be modified to 25 to avoid the overhead of the context switch [1] as shown in equation 2.

$$TQ = \begin{cases} M, & \text{if } M >= 25 \\ 25, & \text{if } M < 25 \end{cases} \qquad (2)$$

### *B. Experimental Frame Work*

The experiment consists of several inputs and outputs attributes. The input attributes consist of TQ, number of processes, CPU burst time and arrival time. The output attributes consist of average turnaround time, average waiting time and number of context switches.

### *C. Data Set*

To evaluate the proposed method, we will take a group of four processes in five different cases with random burst time and random arrival time.

### *D. Performance Metrics*

The proposed algorithm is designed to meet all scheduling criteria such as maximum CPU utilization, maximum throughput, minimum turnaround time, minimum waiting time and minimum context switches. We are considering three performance metrics in each case of our experiment.

- Turnaround Time (TAT)

$$TAT = Finish\ Time - Arrival\ Time \qquad (3)$$

  Average turnaround time should be less.

- Waiting Time (WT)

$$WT = Start\ Time - Arrival\ Time \qquad (4)$$

  Average waiting time should be less.

- Number of Context Switches (CS)

  The number of context switches should be less.

### *E. Experiments Performed*

In each case we will compare the result of the proposed method with Round Robin scheduling algorithm. For RR algorithm, we have taken 20 as the fixed or static TQ.

- Case 1: Assume four processes arrived at time = 0, with burst time (P1 = 12, P2 = 45, P3 = 78, P4 = 90) as shown in TABLE 1. TABLE 2 shows the comparison between RR and MMRR. Figure 3 and Figure 4 shows the gantt chart of RR and MMRR.

TABLE 1. Processes with Burst Time

| Process | Arrival Time | Burst Time |
|---|---|---|
| P1 | 0 | 12 |
| P2 | 0 | 45 |
| P3 | 0 | 78 |
| P4 | 0 | 90 |

TABLE 2. Comparison of RR and MMRR

| Algorithm | Time Quantum | Turnaround Time | Waiting Time | Context Switch |
|---|---|---|---|---|
| RR | 20 | 142.25 | 86 | 12 |
| MMRR | 78,25 | 107.25 | 51 | 4 |

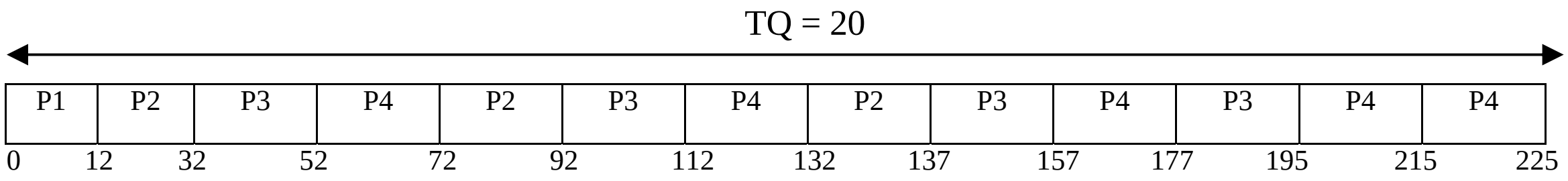

Figure 3. Gantt of RR

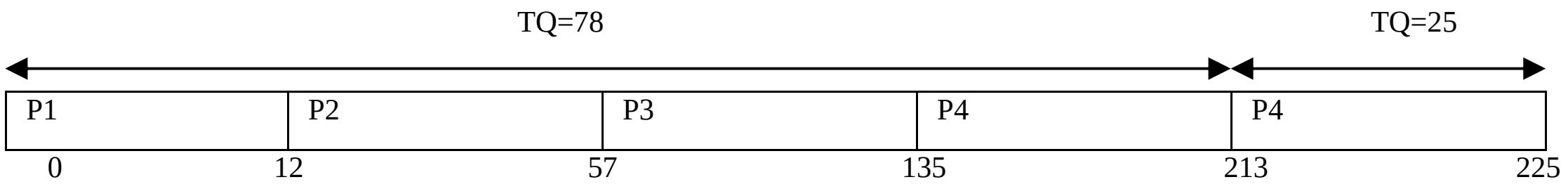

Figure 4. Gantt of MMRR

- Case 2: Assume four processes arrived at time = 0, with burst time (P1 = 61, P2 = 62, P3 = 63, P4 = 64) as shown in TABLE 3. TABLE 4 shows the comparison between RR and MMRR. Figure 5 and Figure 6 shows the gantt chart of RR and MMRR.

TABLE 3. Processes with Burst Time

| Process | Arrival Time | Burst Time |
|---|---|---|
| P1 | 0 | 61 |
| P2 | 0 | 62 |
| P3 | 0 | 63 |
| P4 | 0 | 64 |

.

TABLE 4. Comparison of RR and MMRR

| Algorithm | Time Quantum | Turnaround Time | Waiting Time | Context Switch |
|---|---|---|---|---|
| RR | 20 | 245 | 182.5 | 15 |
| MMRR | 25,25,25 | 230 | 167.5 | 11 |

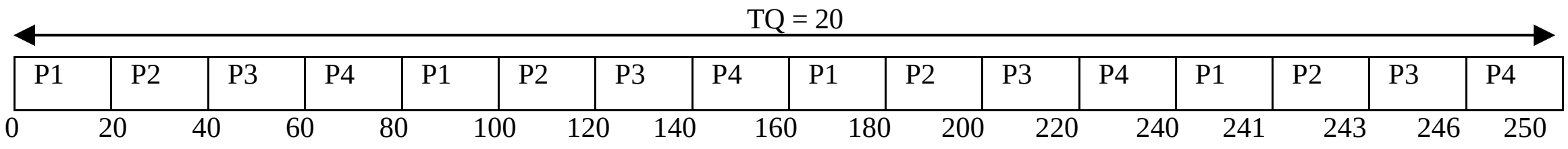

Figure 5. Gantt chart of RR

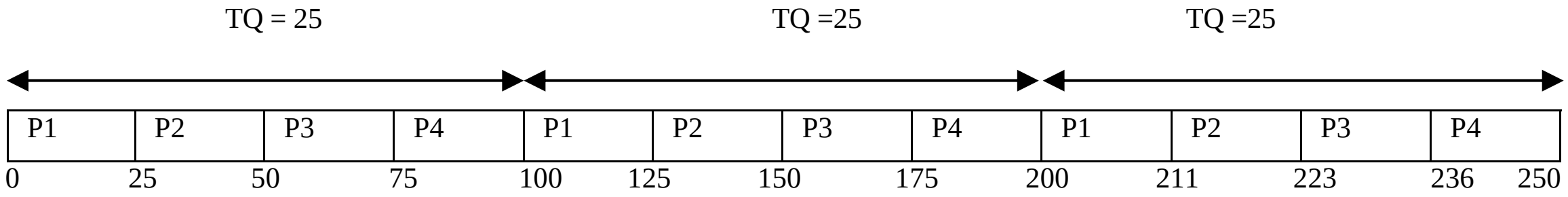

Figure 6. Gantt chart of MMRR

- Case 3: Assume four processes arrived at time = 0, with burst time (P1 = 20, P2 = 40, P3 = 80, P4 = 160) as shown in TABLE 5. TABLE 6 shows the comparison between RR and MMRR. Figure 7 and Figure 8 shows the gantt chart of RR and MMRR.

TABLE 5. Processes with Burst Time

| Process | Arrival Time | Burst Time |
|---|---|---|
| P1 | 0 | 20 |
| P2 | 0 | 40 |
| P3 | 0 | 80 |
| P4 | 0 | 160 |

TABLE 6. Comparison of RR and MMRR

| Algorithm | Time Quantum | Turnaround Time | Waiting Time | Context Switch |
|---|---|---|---|---|
| RR | 20 | 155 | 80 | 13 |
| MMRR | 140,25 | 130 | 55 | 4 |

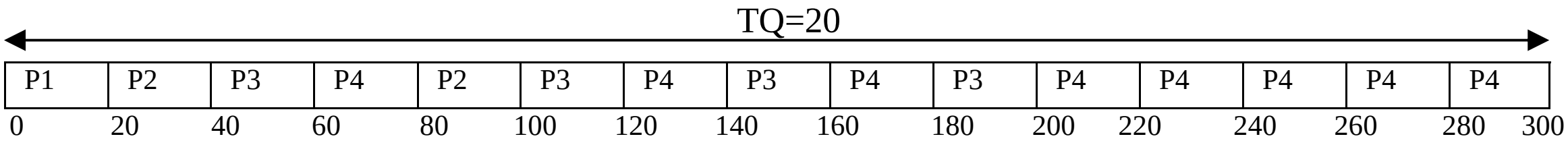

Figure 7. Gantt chart of RR

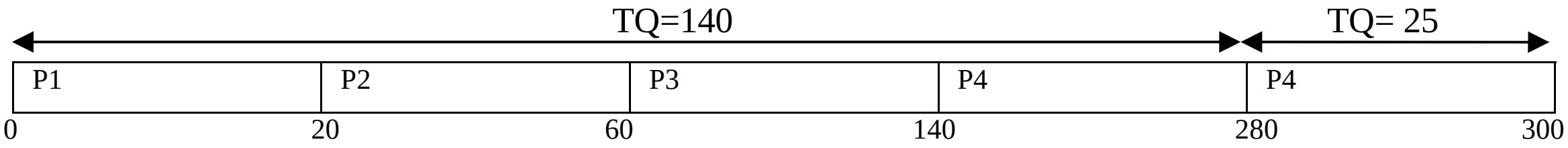

Figure 8. Gantt chart of MMRR

- Case 4: Assume four processes arrived at time (P1 = 0, P2 = 2, P3 = 15, P4 = 23), with burst time (P1 = 5, P2 = 25, P3 = 55, P4 = 75) as shown in TABLE 7. TABLE 8 shows the comparison between RR and MMRR. Figure 9 and Figure 10 shows the gantt chart of RR and MMRR.

TABLE 7. Processes with Burst Time and Arrival Time

| Process | Arrival Time | Burst Time |
|---|---|---|
| P1 | 0 | 5 |
| P2 | 2 | 25 |
| P3 | 15 | 55 |
| P4 | 23 | 75 |

TABLE 8. Comparison of RR and MMRR

| Algorithm | Time Quantum | Turnaround Time | Waiting Time | Context Switch |
|---|---|---|---|---|
| RR | 20 | 80 | 40 | 9 |
| MMRR | 25,25,25,25,25 | 72.5 | 32.5 | 7 |

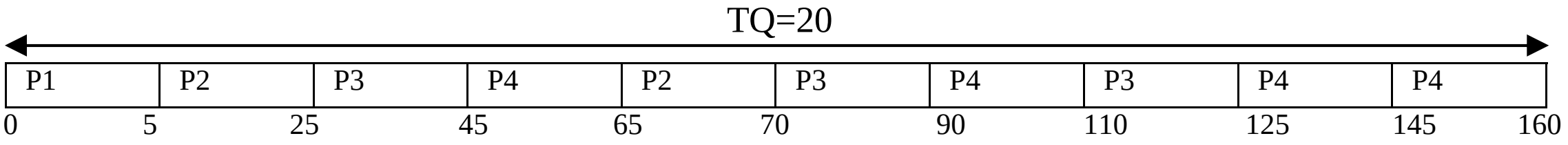

Figure 9. Gantt chart of RR

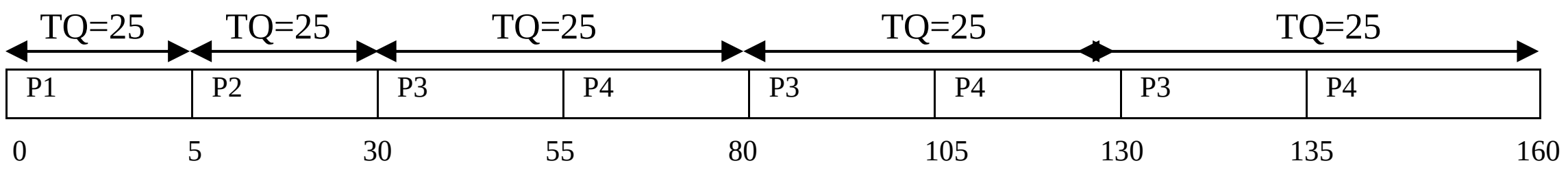

Figure 10. Gantt chart of MMRR

- Case 5: Assume four processes arrived at time (P1=0, P2=17, P3= 35, P4=50), with burst time (P1 = 22, P2 = 47, P3 = 66, P4 = 74) as shown in TABLE 9. TABLE 10 shows the comparison between RR and MMRR. Figure11 and Figure 12 show the gantt chart of RR and MMRR.

TABLE 9. Processes with Burst Time and Arrival Time

| Process | Arrival Time | Burst Time |
|---|---|---|
| P1 | 0 | 22 |
| P2 | 17 | 47 |
| P3 | 35 | 66 |
| P4 | 50 | 74 |

TABLE 10. Comparison of RR and MMRR

| Algorithm | Time Quantum | Turnaround Time | Waiting Time | Context Switch |
|---|---|---|---|---|
| RR | 20 | 133.25 | 81 | 12 |
| MMRR | 25,47,25,25,25 | 95.75 | 43.5 | 7 |

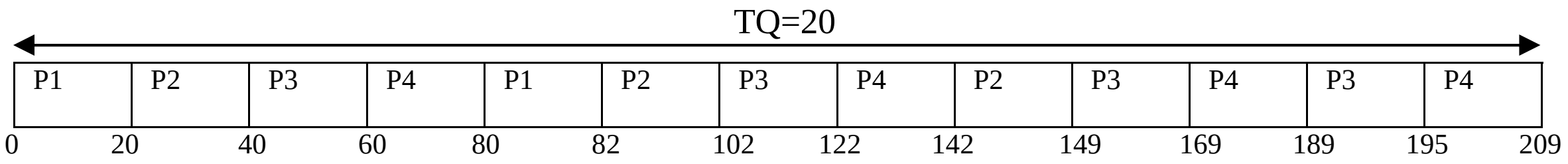

Figure 11. Gantt chart of RR

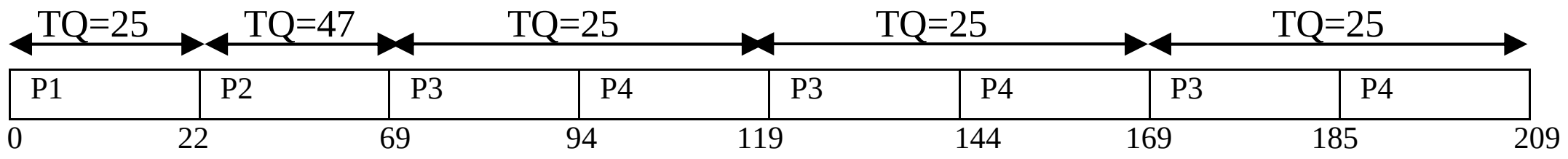

Figure 12. Gantt chart of MMRR

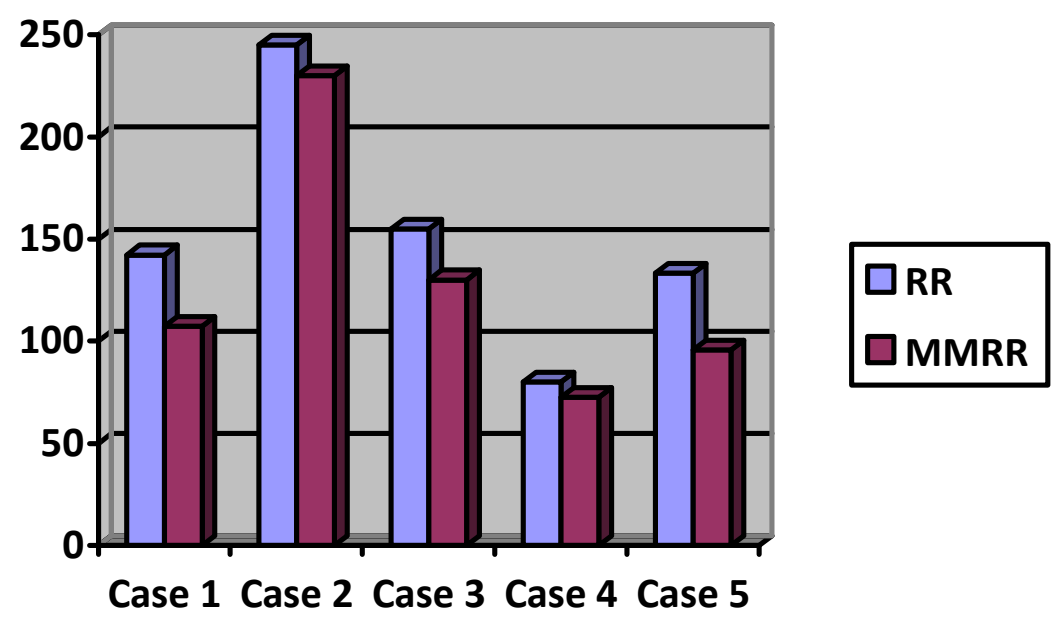

Fig.13: Comparison of by Turnaround Time taking static and dynamic time quantum for case1, case 2, case3 ,case 4 and case 5.

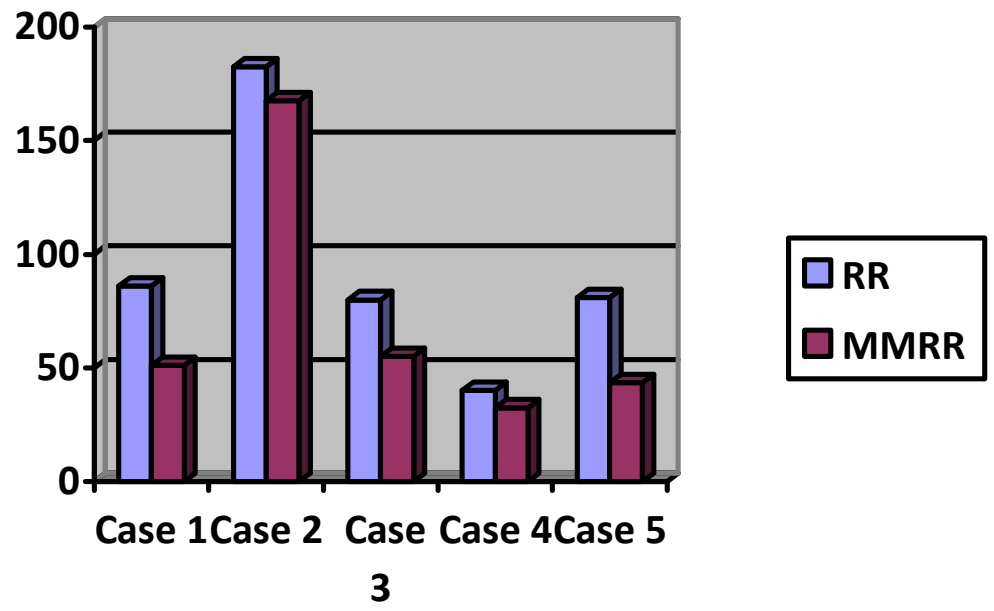

Fig.14: Comparison of Waiting Time by taking static and dynamic time quantum for case1, case 2, case3 and case 4.

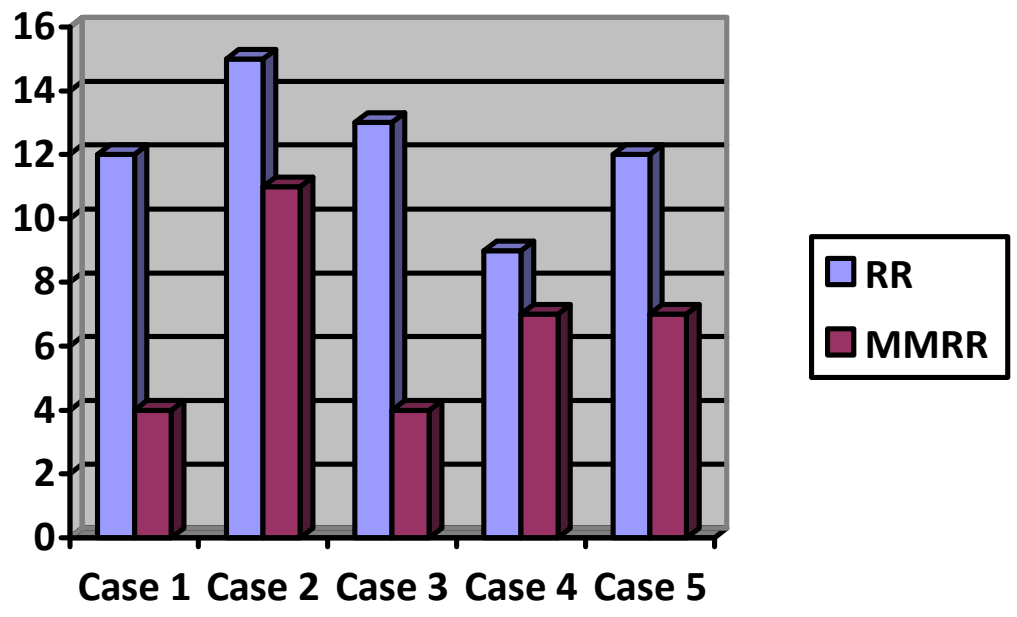

Fig.15: Comparison of Context switches by taking static and dynamic time quantum for case1, case 2, case3 ,case 4 and case 4 .

## REFERENCES

[1] R. J. Matarneh, "Seif-Adjustment Time Quantum in Round Robin Algorithm Depending on Burst Time of the Now Running Proceses", American Journal of Applied Sciences 6 (10), pp. 1831-1837, 2009.

[2] J. Nieh, C. Vaill and H. Zhong, "Virtual-Time Round-Robin: An O(1) Proportional Share Scheduler", Proceedings of the USENIX Annual Technical Conference, Boston, Massachusetts, USA, pp. 25-30, June 2001.

[3] S. M. Mostafa, S. Z. Rida and S. H. Hamad, "Finding Time Quantum of Round Robin CPU Scheduling Algorithm in General Computing Systems using Integer Programming", IJRRAS 5 (1), pp.64-71, October 2010.

[4] A. Silberschatz, P. B. Galvin and G. Gagne, "Operating System Principles", 7th Edn., John Wiley and Sons, 2008.

[5] H. S. Behera, R. Mohanty, and D. Nayak, "A New Proposed Dynamic Quantum with Re-Adjusted Round Robin Scheduling Algorithm and Its Performance Analysis," vol. 5, no. 5, pp. 10-15, August 2010.

[6] A. Bhunia, "Enhancing the Performance of Feedback Scheduling", IJCA, vol. 18, no. 4, pp. 11-16, March 2011.

## AUTHORS PROFILE

Sanjaya Kumar Panda has received his B. Tech degree in Computer Science & Engineering from Veer Surendra Sai University of Technology, Odisha in 2011 and currently pursuing M. Tech degree in Computer Science & Engineering at National Institute of Technology, Rourkela, Odisha.

Sourav Kumar Bhoi has received his B. Tech degree in Computer Science & Engineering from Veer Surendra Sai University of Technology, Odisha in 2011 and currently pursuing M. Tech degree in Computer Science & Engineering at National Institute of Technology, Rourkela, Odisha.